# Developing Excel Thought Leadership

David Lyford-Smith
c/o ICAEW, Chartered Accountants' Hall
1 Moorgate Place
EC2R 6EA
david.lyford-smith@icaew.com

**ABSTRACT**

Over a period of five years, the Institute of Chartered Accountants in England and Wales (ICAEW) has developed a suite of three 'thought leadership' papers surrounding good practice in spreadsheet use and spreadsheet work environments. We will review the history of these three papers, the key lessons which each has to teach, and discuss how the process of making them has helped ICAEW to develop its position in the field.

## 1 TWENTY PRINCIPLES FOR GOOD SPREADSHEET PRACTICE

### 1.1 History

ICAEW was established by royal charter in 1880, combining four previous regional accountancy bodies within the UK. Today, ICAEW is a professional membership organisation and regulator, with over 150,000 members worldwide. ICAEW offers a range of accounting qualifications, supports its members throughout their careers, regulates the professional activities of accountants, and produces thought leadership and technical policy guidance in appropriate fields.

Having established an IT Faculty in the early 1990s to advise on key IT matters for accountants, ICAEW noticed that Excel- and spreadsheet-related content was consistently some of the most popular that they produced. As a result, the spreadsheet-related content was spun off into a separate 'Excel Community' in 2011. In early 2013, a volunteer advisory group was set up, seeking to get spreadsheet experts' input and opinions on activities for the community to pursue.

Starting with its first committee meeting in 2013, this group – the Excel Community Advisory Committee (ECAC) – consisted of a variety of spreadsheet experts from the fields of financial modelling, training, academia, and more (the author was a volunteer member at this time). The group met regularly to discuss both the training and other benefits that the Excel Community made available to its subscribers, but also ideas for public-benefit thought leadership work, under ICAEW's royal charter obligations.

An early theme in the discussions was that of spreadsheet risk. Many contributors felt as though spreadsheet risk was a serious issue affecting world trade, with misuse of spreadsheets being common and a mainly self-taught user-base figuring things out for themselves. But it seemed that little attention was being paid to these issues. So a decision was made to try and tackle the issue by producing a guide to spreadsheet "good practice".

The group individually tried to come up with their own high-level rules for good practice. The format of these attempts differed significantly – some opted for longer lists, some for shorter; some emphasised specific functions and detailed guidance, while others were more principle-driven; some were generalised, and some were divided into separate recommendations for different user groups or applications.

From this list, ICAEW staff member Paul Booth worked to check for commonalities between the authors, and to trim out recommendations that were uncommon or where authors disagreed. Through this analysis, eventually the hundreds of ideas submitted were pared back to a list of twenty that the group could agree upon. These were published in 2014 as the Twenty Principles for Good Spreadsheet Practice[ICAEW, 2014], now into its third edition.

**1.2 Overview**

The Twenty Principles for Good Spreadsheet Practice (20P) are twenty high-level guidance statements for spreadsheet users. The list covers general rules for building a good spreadsheet working environment as well as specific rules about common risky practices.

Example Principles include:
> 2. Adopt a standard for your organisation and stick to it
>
> 14. Never embed in a formula anything that might change or need to be changed.
>
> 19. Build in checks, controls and alerts from the outset and during the course of spreadsheet design.

The 20P are designed to be high-level, such that they are applicable to a range of spreadsheet users in different fields and of different experience levels. They are also principles rather than rules; in some circumstances deviation might be appropriate.

As well as the 20P document itself, ICAEW produced a series of explanatory blogs, webinars, and related content. A facility was made available for organisations to have their financial modelling standards, templates, training courses, or other similar spreadsheet materials ratified as "Twenty Principles Compliant". Several notable standards achieved recognition, including the FAST Standard [FAST Standard Organisation, 2016].

The 20P is designed to start a conversation around best practice, and provide some simple and common-sense recommendations that any spreadsheet user can understand and adopt. It is predicated on the belief that the high-level principles of good practice are industry- and user experience-agnostic, and that they can be applied everywhere.

**1.3 Lessons learned**

The 20P was ICAEW's first experience in creating spreadsheet-focused thought leadership. It was created by overlapping the suggestions of a disparate set of spreadsheet experts and looking for commonalities. This method helped to find the core of agreed good practice from many differing opinions, and would be used repeatedly within ICAEW's later work.

The 20P themselves do not contain any revelatory new ideas about spreadsheet risk and best practice – most of the recommendations are common-sense and even commonplace. However, having a single summary covering a variety of aspects that is not tied to one use case or to a certain class of users is novel and valuable.

As previously pointed out at EuSpRIG, no practice is universally beneficial and worthy of the title of "best practice" [David Colver, 2010]. This was very much in the minds of the committee when trying to identify the most broad recommendations possible – and the choice of "good practice" for the title rather than "best practice" was very much intentional.

The work of Ray Panko and others in the field [e.g. Panko, 2016] focuses on both the need for improvements to spreadsheets – given their high error rates – and the difficulty in using practice recommendations alone to weed out those errors. Panko recommends spending significant effort in team-based code reviews of spreadsheets to improve error rates. Conducting peer review and testing are both principles within the 20P; the paper recommends a broader view of a spreadsheet and the business environment within which it sits rather than just focusing on in-workbook recommendations.

The other work of the Excel Community has been based on the 20P as a founding document.

## 2 SPREADSHEET COMPETENCY FRAMEWORK

### 2.1 History

Following on from the success of the 20P, the ECAC was looking for another area of difficulty with spreadsheets that they could tackle. The subject of spreadsheet capability and knowledge arose as an area of interest – namely, the issues with terms like "spreadsheet expert" or "super user", and the wealth of CVs which use phrases like "proficient with Microsoft Excel". All of these terms are undefined and their use is quite arbitrary – different people will have different ideas of what each means. Furthermore, those with lower ability may lack the metacognitive awareness to be able to accurately assess their own ability. This leads to several problems: Recruiters can't rely on candidates' professed abilities; job applicants can't distinguish themselves as true experts; and trainers have a hard time accurately describing the level of user that their courses are aimed at.

While there are some systems out there for assessing spreadsheet knowledge, the ECAC felt that these were largely either too general – such as the European Computer Driving Licence [ECDL Foundation, 2019] – or which were only concerned with the very top percentile of users, such as Microsoft's own MVP programme [Microsoft, 2019]. Syllabi for training courses often teach to an implicitly identified "appropriate" level, but divining this is not always straightforward.

Based on experience with crafting the 20P, the ECAC set up a working group, the members of which then each submitted their own ideas of what strata existing in spreadsheet ability, and what the key defining knowledge for each level was (the author transitioned to ICAEW staff during this process). After much iteration, the group eventually finalised its work under the title Spreadsheet Competency Framework [ICAEW, 2016], now in a second edition.

**2.2 Overview**

The Spreadsheet Competency Framework (SCF) lays out a set of four categories for spreadsheet users' abilities: Basic User, General User, Creator, and Developer. These four levels are defined by which spreadsheet skills a user of that level would be expected to have, and also what role a person with those skills might play within an organisation.

The four levels are briefly explained as follows:

> A **Basic User** will mostly be carrying out data entry tasks. They will have a grounding in the essential skills needed to avoid major wasted effort or bad practice, but few technical skills beyond that. Anyone that uses spreadsheets should be at this level at a bare minimum.
>
> **General Users** make up the majority of spreadsheet users, essentially modifying and updating spreadsheets on a regular basis. They may have some formula and other more technical knowledge, but are rarely called upon to undertake highly complex tasks, or to make spreadsheets entirely from scratch.
>
> **Creators** use spreadsheets as a core element of their roles, and significantly use its functions and features. They often create spreadsheets from scratch and may create templates and workbooks for users at the first two levels.
>
> **Developers** are the true masters of spreadsheets, with a grasp on the majority of the features of the package and ability to handle many complex tasks. They may be specialists such as modellers, VBA programmers, or statisticians.

The SCF is designed to be of use to anyone needing to refer to spreadsheet ability in a structured and consistent way, although it is written particularly for an accounting and finance audience. Each level has a set of competencies which are 'required' for that level, and others classified as 'beneficial'.

Since publication the SCF has been the subject of a EuSpRIG paper in its own right [Csernoch &Biró, 2017]. This paper consisted of a critique of the SCF and a reworking of the content in the educational sphere.

ICAEW has been exploring avenues to fund and develop an assessment tool that jobseekers, employers, and others could use to place themselves in the appropriate SCF level, allowing them to verify as well as communicate about spreadsheet ability.

**2.3 Lessons learned**

During the creation of the SCF, there was significant disagreement on several issues that identified just how complex of a question this is. For example, the number of levels that existed, and whether a single scale was appropriate for all user groups, was discussed at some length. There is a need to balance precision of measurement with avoiding over-complexity. The eventual compromise was reached of the four described levels, plus a *de facto* fifth level representing those with below-minimum spreadsheet package knowledge that the guide recommends are not permitted to work on spreadsheets until they are given a minimum level of training.

Even once the four-level model was generally accepted by the ECAC team, the mix of core skills at each level was considerably complex. The discussion highlighted just how varied the experience and skills of even a population of expert users could be. For example, some contributors were expert financial modellers with a decade or more of experience, but had never had cause to create a PivotTable. Others were experienced trainers, but did not work much in financial circles, and so might not know how to model a simple loan. VBA programming was also contentious, with some avoiding its use or even arguing that a little knowledge of the subject could be detrimental! The decision to mix between 'required' and 'beneficial' skills in the definition of each level is a nod to trying to acknowledge the possibility of specialisation, even within very expert users.

## 3 FINANCIAL MODELLING CODE

### 3.1 History

Shortly after releasing the 20P, ICAEW ran a roundtable at a ModelOff Global Training Camp event to discuss spreadsheet standards, principles, and best practice, comparing the approach of the 20P with some specific financial modelling standards. Resulting from this discussion, the idea of ICAEW creating a central, non-commercial guide to best practice was first raised. While this idea was of interest, it was not pursued at the time due to resource constraints.

Some years later, a group of expert financial modellers, working outside of ICAEW but including many active volunteers from the ECAC group, began to meet and discuss a joint project to create such a document – a shared vision of what good financial modelling practice looked like. This group began by lining up seven methodologies from various of its members, and looking for commonalities shared between all of them.

Eventually, this discussion generated a rough list of shared ideas. By this time, the external group had agreed that the final product would be most useful if it were finalised and owned by a neutral party external to the group, and identified ICAEW as the preferred destination. Not only is ICAEW seen as a neutral third party, but the work done in the first two publications had helped to establish ICAEW's reputation and expertise in the field.

In early 2018, the early draft of the list was transferred to ICAEW. The ECAC formed a working group, comprised of financial modellers both who had worked on the project at the initiation stage and others, who then reviewed and revised the document as it was reshaped from a list of principles into a more fleshed-out document. The document was also circulated among a wide variety of external stakeholders via an open consultation period, with feedback being actively sought from as many financial modelling experts and organisations as could be identified.

The resulting document was released under the name Financial Modelling Code [ICAEW, 2018], and forms the final part of the trilogy of 'core spreadsheet thought leadership' made by ICAEW.

### 3.2 Overview

The Financial Modelling Code (FMC) is comprised of the essence of seven different modelling methodologies, with review from representatives of around twenty different modelling organisations of different sizes and specialisations. It has been endorsed by a wide range of players in the marketplace, including Microsoft itself.

The FMC consists of a series of recommendations on best practice for tackling common financial modelling problems, which are closely tied in to and cross-referenced with the 20P.  Alongside the guiding principles in each case, a series 'advocated' and 'discouraged' approaches are listed as a suggestion for how one might go about putting the principle into practice.  By intention, few absolute rules are permitted – allowing each modeller to make their own judgment about what's right for them and their clients. Here's an example of a principle from the document:

> **INCLUDE USER GUIDANCE**
>
> (cross-referenced to 20P #7, "Include an 'About' or' Welcome' worksheet to document the spreadsheet")
>
> Although models should be built to require minimal external explanation, appropriate guidance to help facilitate understanding and proper use should be included. This should also include any key assumptions made in calculations (eg, 'all cash flows occur at the end of the appropriate period').
>
> Advocated approaches:
> - Include in the model a worksheet dedicated to being a user guide.
> - Embed a separate user guide document in the workbook.
> - Add contextual user guidance throughout the model where appropriate.
>
> Discouraged approach:
> - Don't store documentation in a separate file or email that may become separated from the model itself.

The FMC taken as a whole is a guidance document, and not a how-to guide – there are no Excel screenshots in the document and no formulas.  There are other publications in the marketplace aimed at being modelling guides, many of which discuss best practice, but the FMC is aiming to be a higher-level guidance document than these, or even than modelling standards such as FAST.

An option for organisations to become official supporters of the FMC is available; at time of writing, eleven such organisations have put themselves forward, including several 'Top 10' accounting firms and Microsoft themselves.

### 3.3 Lessons learned

This was the first piece of work that ICAEW produced that originated outside of the organisation.  While the final piece underwent considerable iteration and was reviewed by many sources, the original work was done by a separate group.  However, the methodology used was similar to ICAEW's previous two spreadsheet works – a group of domain experts was convened, and their knowledge was combined and contrasted to settle on a core of agreed practice.

Opinions about the best way to perform financial modelling are remarkably divergent.  To take a clear example, opinions on the use of named ranges runs the gamut from "essentially forbidden" to "essentially mandatory".  To create a set of principles that would accept both extremes but not favour a particular solution required careful wording and consideration.  In general, the final document comments on the aims of good practice more than how to achieve them in the main text, and several (sometimes mutually exclusive) options of how to achieve those aims are included.

While the SCF did include reference to the 20P, the creation of the FMC leant more heavily on the 20P, and the final version includes both a list of the Principles and frequent cross-references to them. While the basic principles were developed by the external expert group, the resulting output closely aligned to the 20P and finding connections was straightforward. This shows that the 20P truly represent well-understood guiding principles of good practice.

**4 CONCLUSION**

Over the period since the ECAC was formed in 2013, ICAEW has worked on a variety of spreadsheet-related projects. The common methodology for the three major thought leadership pieces was, in brief:

1. Identify an area where a high-level guidance publication could be of use by consulting with experts and stakeholders
2. Convene a suitable and diverse group of experts and synthesise a consensus view from their disparate opinions
3. Invite commentary and critique via an exposure draft process
4. Complete the project based on the feedback received

This process has been successful in producing a trilogy of work which is of a high standard and which is well-regarded and accepted. Nevertheless, it is interesting that all three writing processes uncovered a wealth of individual preference, difference of opinion, and variety in what "good" really meant in the realm of spreadsheets. No doubt one of the difficulties that has plagued attempts to reduce spreadsheet risk in practice is this very heterogeneous field of viewpoints. But spreadsheet use is also very diverse, and perhaps it is only natural that agreement can be hard to find.

All three projects had to tackle similar issues with forging common ground out of many differing views. Spreadsheets are in use across a very broad cross-section of business and other life, and naturally users will have different priorities, experiences, and opinions. Our approach in each case was to identify the most broadly agreeable version of contentious statements, making things more generic or more fundamental where possible. Leaving room for individual users to interpret and enact our recommendations in their own way was an active decision.

Spreadsheet risk remains a significant issue in modern business, and stories of spreadsheet mistakes tripping up companies continue to be common. So it's reasonable to ask how the cause of fighting spreadsheet risk has been affected by ICAEW's work in this area, and what remains to be done. These three guides have all been well-received, but the key issue remains attracting attention to these issues and convincing the audience to seek out recommendations for improving their practice. Anecdotally, many users have been happy to take on recommendations for improvements when they have been presented with them, but do not actively seek them out. ICAEW's strategy is to continue to promote both the case for improving practice, and its trio of publications aimed at meeting that end.